\documentstyle[sprocl]{article}

\def\PRD{{\em Phys. Rev.} D}
\def\ZPC{{\em Z. Phys.} C}

\def\ra{\rightarrow}

\def\be{\begin{equation}}
\def\ee{\end{equation}}
\def\bea{\begin{eqnarray}}
\def\eea{\end{eqnarray}}

\newcommand{\ba}{\begin{eqnarray}}
\newcommand{\ea}{\end{eqnarray}}
\newcommand{\no}{\nonumber}

\begin{document}
\begin{flushright}
IFT 97/7 \\
June 1997\\
{\bf hep-ph/9708274}
\end{flushright}
\vskip 0.5cm
\title
{PROBING THE STRUCTURE OF VIRTUAL PHOTON IN THE DEEP INELASTIC
COMPTON PROCESS AT HERA}

\author{ MARIA KRAWCZYK\thanks{}, A. ZEMBRZUSKI}

\address{Institute of Theoretical Physics, University of Warsaw, 
ul. Hoza 69,\\ 00-681 Warsaw, Poland}


\footnotetext{$^a$ contribution to PHOTON'97, 10-15 May 1997,
Egmond aan Zee, The Netherlands}

\maketitle\abstracts{
The sensitivity of the  Deep Inelastic Compton (DIC) 
scattering at HERA 
to the structure of the virtual photon is discussed.
It is demonstrated that  the gluonic content of the
virtual photon can be pinned down by 
measuring the   photons with $p_T \sim 5 $ GeV in the 
proton direction. 
}

\section{Introduction}
The Deep Inelastic Compton (DIC)  process provides 
the opportunity to probe at HERA the structure of the 
real photon, as 
 was  pointed out in   \cite{mk,bks,prob,aurenche,gs}
\footnote { 
The sensitivity to  the quark fragmentation 
into the photon in the DIC process has been studied 
as well, see \cite{bks}.}. 
 The ability to probe the structure of the photon 
(and the proton) in the tagged and untagged events  at HERA
was studied in paper \cite{prob}  
with the conclusion that in order to separate the 
contribution arising from
the gluonic content of the (real) photon the tagged 
condition is preferred.

The first attempt to describe the DIC process at HERA using the 
structure of the virtual photon can be found in \cite{za,azem},
where the EPA approach was compared with the  calculation, 
where the virtual photon interacts directly or by its partonic 
content.
Results obtained for the virtual photon were based 
on the naive Parton Model formulae {only for quarks}. 
 The EPA approach
 leads to  the $p_T$ distribution in the resolved photon
 subprocesses higher   by about 20-45 \%  than one based on the 
 "exact" approach using the virtual photon structure.
 For the subprocesses due to the direct photon interaction
 the  EPA works much better.

In the present  paper  we examine,  
using the   GRS (LO) \cite{grsparam} parton parametrization
for the virtual photon, the usefulness of DIC process  
to study at HERA the structure of a {\underline {virtual photon}},
in particular the gluonic content of virtual photon.   
The recent study have   showed that 
the parton distributions of the virtual photon can be tested
at HERA via tagged single  high $E_T$ jet  or $b \bar b$ production 
\cite{grs}.
\section{Deep Inelastic Compton scattering at HERA collider}
We investigate a inclusive DIC  process in which photons
with large transverse momentum, $p_T\gg\Lambda_{QCD}$, 
are produced in
electron(positron)-proton collision:
\be
ep\ra e\gamma X.
\ee
For relatively small momentum transfers between electrons, 
this reaction proceeds by the 
exchange of a virtual photon, i.e. the $Z$ exchange can be neglected. 
We will limit ourself to events with the tagged electrons,
so the energy and 
(positive) virtuality of the initial photon $-p^2=P^2$ 
can be estimated. 

Depending on the conditions (untagged, antitagged or tagged events) 
DIC process may proceed
via the  (almost) real photon-proton collision:
\be
\gamma p \ra \gamma X
\ee
or via the virtual photon-proton scattering:
\be
\gamma^* p \ra \gamma X.
\ee
The comparison between these two   approaches to DIC 
process at HERA can be found in \cite{za,azem}.
The direct photon interaction for both real and virtual photons, 
corresponding to the following subprocesses
\be
\gamma q_{p}\ra\gamma q,
\ee
\be
\gamma^* q_{p}\ra\gamma q,
\ee
where the initial photon  interacts with a quark from the proton 
(the Born approximation), dominates at $very$ $large$ $p_T$ $\sim 
\sqrt{S_{ep}}/2$.
 
In the $moderate$ $p_T$ region, 
$\Lambda_{QCD}\ll p_T\ll\sqrt{S_{ep}}/2$,
  a resolved photon processes,
\i.e. where the photon interacts through its partonic constituents,
become important.
There are three   types of  subprocesses 
involving the partonic constituents  of the {\underline {initial}} 
and/or {\underline {final}}
photons in DIC (below we discuss only the virtual initial photons):
\begin{itemize}
\item
single resolved initial photon
\be
g_{\gamma^*} q_p\ra\gamma q,
\ee
\be
q_{\gamma^*}g_p\ra\gamma q,
\ee
\be
q_{\gamma^*}\overline q_p\ra\gamma g
\ee
\be
\overline q_{\gamma^*} q_p\ra\gamma g
\ee

\item 
single resolved final photon
(fragmentation into the photon)
\be
\gamma^* g_p\ra q {\bar q} 
\ee
\be
\gamma^* q_p\ra g { q} 
\ee

\item double resolved photons
\be
g_{\gamma^*} g_p\ra g g
\ee
\be
q_{\gamma^*} g_p\ra q g, etc.
\ee
 
\end{itemize}
In this paper we limit ourself to the process 
involving (single) resolved 
initial photon, where we expect to see 
the effect due to the gluonic content in the 
virtual photon  $P^2\neq 0$. The full discussion 
will be given elsewhere
\cite{azem}.
\section{Calculation of the cross section}
The differential cross section for the deep inelastic 
electron-proton scattering
with a photon in the final state, eq. 1, can be written in 
the following way:
\ba
E_eE_{\gamma}{d\sigma^{ep\ra \gamma eX}\over 
d^3p_ed^3p_{\gamma}}
=\Gamma \Bigl (E_{\gamma}{d\sigma^{\gamma^*p\ra 
e\gamma X}\over 
d^3p_{\gamma}}|_T+\epsilon 
E_{\gamma}{d\sigma^{\gamma^*p\ra e\gamma X}\over
d^3p_{\gamma}}|_L\Bigl).
\ea
where $E_e(E_{\gamma})$ and $p_e(p_{\gamma})$ are energy 
and momentum of the final state electron(photon).
Coefficients $\Gamma$ and $\epsilon\Gamma$ 
(functions of energy and momentum of the electron in 
initial and final states; see \cite{collins}) can
be interpreted as the probability of emitting by the initial 
electron a
virtual photon polarized transversely and longitudinally. 

Since the cross section for the 
reaction $ep\ra e\gamma X$ is dominated by 
the exchane of photons with  small virtuality, 
one  can neglect a contribution due to  
the longitudinal polarization (see also \cite{za}).
Assuming that exchanged photons have only
transverse polarization we obtain:
\ba
E_eE_{\gamma}{{d\sigma^{ep\ra e\gamma X}}
\over{d^3p_ed^3p_{\gamma }}}
=\Gamma E_{\gamma}{{d\sigma^{\gamma^*p\ra e\gamma X}}\over 
{d^3p_{\gamma}}}|_T.
\ea
Performing in the above cross section the 
integration over the 4-momentum of the
final electron one
obtain the differential cross section
$E_{\gamma}{d\sigma^{ep\ra \gamma X}\over d^3p_{\gamma}}$ 
\ba
E_{\gamma}{d\sigma^{ep\ra \gamma X}\over d^3p_{\gamma}}=
\int {d^3p_{e}\over E_{e}}
\Gamma
E_{\gamma}{d\sigma^{\gamma^*p\ra\gamma X}\over d^3p_{\gamma}}.
\ea
where the 
flux can be found in \cite{collins,azem} and the 
invariant cross section 
$E_{\gamma}{d\sigma^{\gamma^*p\ra\gamma X}\over 
d^3p_{\gamma}}$ has a form:
\begin{itemize}
\item for the direct (Born) process:                              
\ba
\Bigl (E_{\gamma}{d\sigma^{\gamma^*p\ra\gamma X}\over 
d^3p_{\gamma}}\Bigr )_{dir}
=\displaystyle\sum_{q,\overline q}\int\limits_0^1dx_p
\ \
f_{q/p}(x_p,\tilde Q^2)
\ \
E_{\gamma}
{d{\hat {\sigma}}^{\gamma^*q\ra\gamma q_p}\over d^3p_{\gamma}},
\ea
\item for processes involving the resolved initial photon:
\ba
\Bigl (E_{\gamma}{d\sigma^{\gamma^*p\ra\gamma X}\over d^3p_{\gamma}}
\Bigr )_{res}   \\ \no
=\displaystyle\sum_{i,j}\int\limits_0^1dx_{\gamma}
\int\limits_0^1dx_p
f_{i/{\gamma^*}}(x_{\gamma},\tilde Q^2,P^2)
\ \
f_{j/p}(x_p,\tilde Q^2)
\ \
\Bigl (E_{\gamma}{d{\hat {\sigma}}^{i j\ra\gamma k}\over 
d^3p_{\gamma}}\Bigr ),
\ea
\end{itemize}
$f_{i/\gamma^*}(x_{\gamma},\tilde Q^2,P^2)$ 
is a ($i$) parton distribution
in the virtual photon with a virtuality equal to $P^2$ 
at scale $\tilde Q^2$. The cross sections $\hat \sigma$ 
correspond to the
partonic subprocesses.

In the calculation  we take into consideration the
virtuality (-$P^2$) of the photon emitted by 
the electron as it follows from the kinematics of the process. 
\section{Results}
In the calculation we used 
the GRS (LO) parton parametrizations  for the parton  
distributions 
in the  virtual photon \cite{grsparam} and
the GRV (LO) set of the quark and the gluon demsities
for the proton \cite{grv}. 

Calculations were performed for the 
energy of the HERA accelerator: $S_{ep}=98400$
GeV$^2$. We assumed the number of flavours $f=4$, 
the QCD parameter 
$\Lambda_{QCD}=0.2$ GeV,
and the energy scale  in Eqs. 18-19 
equal to the transverse momentum of the final photon: 
$\tilde Q=p_T$.

We calculated the cross section for the transverse momentum 
of the final photon equal to 5 GeV and for the fixed energy 
of the initial photon:  $E_{\gamma}=0.5 E_e$ 
(so the $y$ variable was 
equal to 0.5).  The rapidity dependence 
was studied for the various values
of $P^2$ between 0.03 to 2.5 GeV$^2$. \footnote {Note that  
tagging of the final electron
helps to distinguish the direct contribution 
from the resolved ones.}

The results for the 
$E_{\gamma}{d\sigma^{ep\ra e\gamma X}\over 
{d^3p_{\gamma} dP^2 dy}}$ 
for the subprocesses Eqs. 5 and 6 for three values of $P^2$ 
are presented 
in Fig.1a as a function of the rapidity
\footnote{The rapidity Y is equal to 
$Y=-\ln tg{\theta\over 2}$, where
$\theta$ is the angle between the momentum of the photon in the 
final state and the
momentum of the electron in the
initial state.} in the electron-proton center of mass system.
In Fig.1b the comparison of the different contributions 
to the considered cross section is plotted 
for the $P^2=0.25$ GeV$^2$. 
\vskip 0.5cm
\begin{figure}[ht]
\vskip 2.1in\relax\noindent\hskip -1.50in
       \relax{\includegraphics{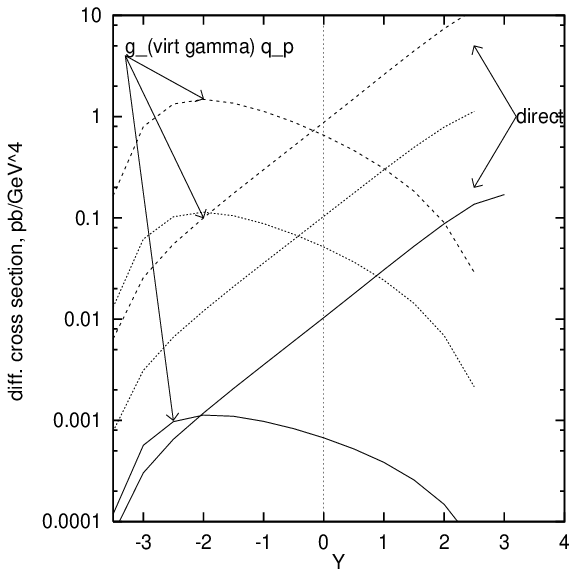}}
	\relax\noindent\hskip 3in
       \relax{\includegraphics{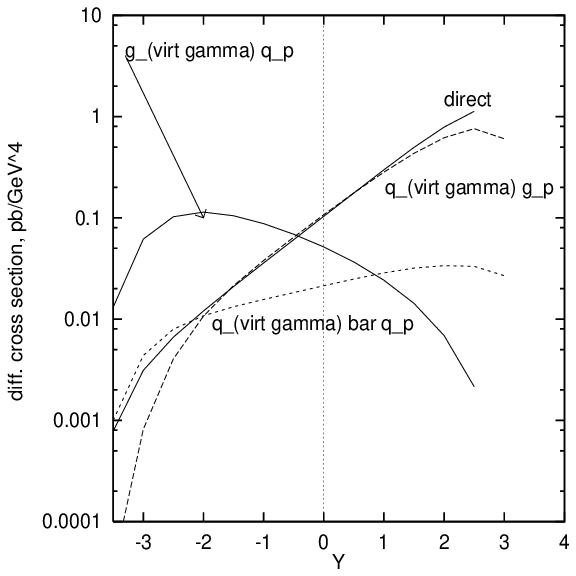}}
\caption{ {a) The cross section
$E_{\gamma}{d\sigma^{ep\ra e\gamma X}\over 
{d^3p_{\gamma} dP^2 dy}}$ 
for subprocesses: direct (5) and $g_{\gamma^*}q_p\ra
\gamma q$~(6) at the $P^2$=0.03,~0.25 and 2.5 GeV$^2$
(dotted, dashed and solid line). b) The same for 
subprocesses: (5) and (6) (solid lines), 
(7) (dashed line) and (8+9) (dotted line)~at~$P^2$=0.25~GeV$^2$.
 }}
\label{fig:excl}
\end{figure}
The clear dominace of the contribution due to 
gluonic content of the virtual photon 
in the proton direction is seen for the 
considered range of virtuality
of initial photon. We check that this holds also for smaller $P^2$
values. 
\footnote{Due to smooth behaviour of the GRS parametrization  
in the limit $P^2\ra 0$ we were able
to  perform the calculation also below $\Lambda^2_{QCD}$.} 

The discussion based on the full set of diagrams, wider $P^2$ range
and others parton parametrizations will be given elesewhere\cite{az}. 
Note that    
the interference with the Bethe-Heitler process, discussed in 
\cite{brodsky}, seems to be  small 
for $p_T$=5GeV and
in the region of the rapidity where the 
gluonic content of the virtual photon plays 
a dominante role\cite{pj}.
\section{Conclusions}
Tagged DIC events at HERA were studied 
using the GRS and GRV parton
parametrizations for the direct and resolved
virtual photon subprocesses. 
The  contribution due to gluonic content of the virtual photon
was found to dominate in the direction of the proton 
as compared to others subprocesses. 
This can have important consequences for the possibility 
of measuring 
the gluon content of the virtual photon at HERA.

\section*{Acknowledgments}
One of us (MK) would like to thank organizers of the very fruitful 
conference.  
We are grateful to Stan Brodsky for pointing  the reference 
\cite{brodsky}
and useful discussions. We thank  M. Stratman and A. Vogt
for sending us the Fortran code 
with the parton parametrizations.
Supporting by the Polish Committee for Scientific Research, 
Grant No 2P03B18209.


\section*{References}
\begin{thebibliography}{99}
\bibitem{mk}
M.~ Krawczyk,
 Acta Physica Pol. {\bf B21} (1990) 999.
\bibitem{bks}
A.~C.~Bawa, M.~Krawczyk, W.~J.~Stirling,
Z.\ Phys. {\bf C50} (1991) 293.
\bibitem{prob} A.~C.~Bawa and M.~Krawczyk,
Probing the structure of proton and photon
in deep inelastic Compton process at HERA and LEP/LHC,
 Proc. ``Physics at HERA'',  Hamburg 1991, vol.
1, p. 579, and  IFT 91/17.
\bibitem{aurenche}
P. Aurenche, et al. , proceeding of the HERA Workshop, 
Hamburg 1987, p. 561 
and \ZPC 56 (1992) 589
\bibitem{gs}
L. E. Gordon, J. K. Storrow, \ZPC 63 (1994) 581
\bibitem{za}
A.~ Zembrzuski and  M.~Krawczyk,
On the validity of the equivalent photon approximation 
and the structure of a virtual photon,
 Proc. ``Physics at HERA'', Hamburg 1991, vol.
1, p. 617, and Warsaw Univ. IFT 91/15.
\bibitem{azem} A. Zembrzuski, The $ep\ra e \gamma X$ 
process at HERA.
Structure of photon, Msc Thesis, Warsaw University 1991 
\bibitem{az} M.~Krawczyk and A.~ Zembrzuski, in preparation    
\bibitem{grsparam}
M. Gl\"{u}ck, E. Reya and M. Stratmann, \PRD 51 (1995) 3220
\bibitem{grs} M. Gl\"{u}ck, E. Reya and M. Stratmann, 
\PRD 54 (1996) 5515;
D. de Florian, C. Garcia Canal, R.Sassot, 
CERN-TH/96-234 (hep-ph/9608438);
M. Klasen, G. Kramer, B. Plotter, DESY 97-0.39(hep-ph/9703302);
B. Plotter,  DESY 97-138 (hep-ph/9707319)
\bibitem{collins} F. Halzen , A. D. Martin, Quarks and Leptons,
John Wiley \& Sons, 1984
\bibitem{grv} M. Gl\"{u}ck, E. Reya and A. Vogt, 
\ZPC 67 (1995) 433
\bibitem{brodsky} S.J.Brodsky, private communication and
S.J.Brodsky, J. F. Gunion, R. L. Joffe, \PRD 6 (1972) 2487
\bibitem{pj} P. Jankowski and M. Krawczyk, in preparation

\end{thebibliography}
\end{document}